\documentstyle[12pt]{article}

\begin{document}
%July 14, 1995

%\documentstyle{article}
%\topmargin -0.6in
%\textheight 22cm
%\textwidth 15cm
%\Roman{section}
\renewcommand{\thesection}{\Roman{section}}

\baselineskip=24pt
\begin{titlepage}
\hspace{10cm} AS-ITP-95-022
\vspace{1cm}

\centerline{\large \bf ON OBSERVATION OF $e^+e^-\rightarrow b 
\overline{b}Z^0$ AT LEP II}

\vspace{0.2cm}

\centerline{\large \bf WITH TWO HIGGS DOUBLETS}

\vspace{1.0cm}

\centerline{Chao-Hsi Chang $^{1,2}$, Yi-Bing Ding $^{1,3}$, Xue-Qian Li
$^{1,4}$, Jian-Xiong Wang $^{5}$ and Yue-Hong Xie $^{4,5}$}
\vspace{1cm}
{\small
\begin{center}
1.   China Center of Advanced Science and Technology,
  (World Laboratory)

P.O.Box 8730 Beijing 100080, China.
\vspace{4pt}

2. Institute of Theoretical Physics, Academia Sinica,

P.O. Box 2735, Beijing 100080, China\footnotemark[0]{*}\footnotetext[0]
{$^*$Mailing address}.

\vspace{4pt}

3. Department of Physics, Graduate School, Academia Sinica,

Beijing, 100039, China\footnotemark[0]{*}.
\vspace{4pt}

4. Department of Physics, Nankai University,

Tianjin, 300071, China\footnotemark[0]{*}.
\vspace{4pt}

5. Institute of High Energy Physics, P.O. Box 918-4,

Beijing, 100039, China\footnotemark[0]{*}.
\end{center}
}
\vspace{1.5cm}

\centerline{\large \bf Abstract}
\vspace{1cm}

We study possible observational effects of the two Higgs doublets in the
$e^+e^- \rightarrow b \overline{b} Z^0$ at the LEP II energy. We have found 
that the observational values can be obviously different from that predicted 
by the minimal Standard Model (MSM), but the results depend on the parameters
of the extended model. The possibilities of the observation are discussed 
in some details.\\

\end{titlepage}

\baselineskip 22pt

\section{Introduction}

\vspace{0.5cm}

The Standard Model (SM)
has achieved great successes in almost all fields of
phenomenology of high energy physics so far.  Especially,
the top quark mass  has been published as
$176\pm 8({\rm stat.})\pm 10({\rm sys.})$ GeV
\cite{CDF1} and $199^{+19}_{-21}({\rm stat.})\pm 22 ({\rm sys.})$ GeV
\cite{D0}, thus the three generation structure of the SM is complete. The
only still obscure part
in the theory is the Higgs sector,
which is crucial to the mechanism
of the Spontaneous Symmetry Breaking (SSB) of $SU_L(2) \times U_{Y}(1)$.
Therefore, more attention will be focused on the "Higgs hunting"
within a wide energy range \cite{Montagna}.

The Higgs hunting includes two-folds. The first is to seek for the
existence evidence
of Higgs bosons through all available experiments, whereas
the other is to test if the Higgs sector is indeed that of the Minimal
Standard Model or an alternative, for example, an extension
with two or multi Higgs doublets.

There has been much effort to search for the Minimal Standard Model (MSM)
Higgs at lower energies of LEP I, but so far no success has
ever been reported. With the top-quark being
discovered, one cannot elude this acute question now.
LEP II will open a new  place for
Higgs hunting, because
clearer signals for heavier Higgs are expected  above  the relatively
low background.
Recently Boos and Dubinin estimated the Higgs signal at process
$e^+e^- \rightarrow b \overline{b} Z^0$ \cite{Boos} and they found that
the ratio of
the Higgs signal versus the background  may approach to unity as
$\sqrt{s} \sim 200$ GeV provided $m_H \sim 100$ GeV. If the MSM is  right,
namely only one neutral Higgs exists, the situation for
determining the mass of Higgs is optimistic from this estimate on the
suggested measurement.
Because there is no any free parameter
except the Higgs mass at the tree level,
a precise data of the cross section and differential
cross section of
$e^+e^-\rightarrow b\bar bZ^0$ would pin down the Higgs mass almost,
if $M_H\leq 100$GeV/c$^2$. However, if
there are more Higgs doublets, it will be another story.

In fact, search for some mechanisms beyond the Minimal Standard Model (MSM)
is also interesting for both experimentalists and theoretician of high
energy physics \cite{Gunion}.
The "minimal" extended Standard Model(ESM) is the gauge theory
$SU_L(2) \times U_Y(1)$
with an extended Higgs sector consisting of two Higgs doublets \cite{Abbott}.
If the  Higgs sector is indeed more complicated,
they may also play roles in the LEP II,
such as $e^+ e^- \rightarrow b \overline{b} Z^0$,
The neutral
Higgs being real or virtual,
can directly contribute to the process, so it may  provide
us with the information of Higgs sector.

Recently, a more accurate measurements on $B\rightarrow K^*\gamma$
and $B\rightarrow X_s+\gamma$ set an
upper limit to $b\rightarrow s\gamma$ transition and establish
more stringent constraints to all the extended Standard Models on the Higgs
sector.
\cite{stange}, \cite{hewett} and \cite{grant}.
When a special extended Standard Model is applied to the concerned
process, the constraint must be taken care of seriously.

For the cross section evaluation of $e^+e^-\rightarrow b\bar bZ^0$, we will
show below that just because of existence of more neutral Higgs, the situation
becomes much more complicated and one cannot be so optimistic about hunting
Higgs from the data.

In this work, we analyze the contribution of two Higgs doublet model to
$e^+ e^- \rightarrow b \overline{b} Z^0$, it is noted that existence of
the second neutral
Higgs particle $h^0$ can cause an obvious difference
at the differential cross section from that
by $H^0$ only.
This result suggests that if only total cross section of $e^+ e^- \rightarrow
b \overline{b} Z^0$ is measured, the obtained value cannot determine the
Higgs mass unless there is only one Higgs doublet, however it
indeed can if measuring the differential cross section with respect to $(p_2 + p_3)^2
\equiv s_2$ precisely where
$p_2$ and $p_3$ are the momenta of $b$ and $\overline{b}$ respectively.\\

\section{The extended Standard Model with two Higgs doublets}

The general description of the models can be found in ref. \cite{Gunion}.
Here we just give some necessary information to make the paper more
self-content.

There are two types of the model
where the quarks gain masses in different ways and we will denote them as
Model I and Model II as in the literature.

The key parameter is $\beta$ which is defined as
\begin{equation}
tan \beta = v_2/v_1
\end{equation}
where $v_1$ and $v_2$ are the vacuum expectation values (VEV) of the two
Higgs doublets. In model I, quarks and leptons gain masses only from the second
Higgs doublet while the first Higgs doublet decouples. In contrast, in
model II, d-type quarks and leptons obtain masses from the first
doublet whereas u-type quarks from the second doublet. There are three neutral
bosons $H^0, h^0$ and $A^0$ remain as real particles after SSB, but since
$A^0$ is a CP-odd boson, in our case (at the tree level)
it does not contribute at all.

The Lagrangian for Higgs-fermion coupling can read
\begin{eqnarray}
L_{Hf\overline{f}} &=& - \frac{g}{2M_W\sin \beta} \overline{D}M_DD(H^0\sin \alpha
+h^0\cos \alpha)- \frac{ig\cot\beta}{2M_W } \overline{D} M_D \gamma_5 DA^0
\\ \nonumber
 & & - \frac{g}{2M_W \sin \beta} \overline{U} M_U U(H^0 \sin \alpha +
 h^0 \cos \alpha) + \frac{ig \cot \beta}{2M_W} \overline{U} M_U \gamma_5 U A^0
\\ \nonumber
 & & +\frac{g \cos \beta}{2 \sqrt{2} M_W} (H^+\overline{U} [M_UK(1-\gamma_5)
 - KM_D(1+\gamma_5)]D + h.c.)
 \end{eqnarray}
 for Model I. In contrast, the Model II interaction  is
 \begin{eqnarray}
L_{Hf\overline{f}} &=& - \frac{g}{2M_W\cos\beta} \overline{D}M_DD(H^0\cos \alpha
-h^0\sin \alpha)+ \frac{ig\tan \beta}{2M_W } \overline{D} M_D \gamma_5 DA^0
\\ \nonumber
 & & - \frac{g}{2M_W \sin \beta} \overline{U} M_U U(H^0 \sin \alpha +
 h^0 \cos \alpha) + \frac{ig \cot \beta}{2M_W} \overline{U} M_U \gamma_5 U A^0
\\ \nonumber
 & & +\frac{g}{2 \sqrt{2} M_W} (H^+\overline{U} \cot \beta[M_UK(1-\gamma_5)
 + \tan \beta KM_D(1+\gamma_5 )]D + h.c.),
 \end{eqnarray}
where K is the Cabibbo-Kabayashi-Maskawa matrix, $M_U$ and $M_D$ are the mass
matrices of the u-type and d-type quarks, $\alpha$ denotes a mixing
between $H^0$ and $h^0$ as
\begin{equation}
H^0=\sqrt{2} [(Re\phi_1^0 - v_1)\cos\alpha + (Re\phi_2^0 - v_2) \sin\alpha ]
\end{equation}
\begin{equation}
h^0=\sqrt{2} [-(Re\phi_1^0 - v_1)\sin\alpha + (Re\phi_2^0 - v_2) \cos\alpha ].
\end{equation}

Since none  of the parameters $M_{H^0}, M_{h^0}, \beta$ and $\alpha$ is determined
experimentally, the extra Higgs doublet
increases complexity for identifying Higgs and we
will discuss the measurement problem later.\\

\section{The  cross section and differential cross section of
$e^+e^- \rightarrow b\overline{b}Z^0$}

For the MSM,
totally there are nine different Feynman diagrams at the tree level,
which are given in ref.\cite{Boos}, however, in the ESM,
because of existence of h$^0$, the
diagram (1-3) of ref.\cite{Boos} should be split into two diagrams corresponding to
H$^0$ and h$^0$ respectively. Being explicitly, we demonstrate the diagrams in Fig.1
of this paper. The ten diagrams interfere,
so the calculation is tedious but straightforward.
We first write down the amplitude contributed from the ten Feynman
diagrams and  then employ a
standard program for  numerical evaluation of the cross section.

The propagator of H$^0$ and h$^0$ is written as
\begin{equation}
\Delta = \frac{i}{p^2-m_{H}^2+i\Gamma _Hm_H}
\end{equation}
where $\Gamma _H$ and m$_H$ are the mass and width of H$^0$ or h$^0$ respectively.

At the tree level, $\Gamma_{H^0}$, $\Gamma_{h^0}$ in model I and model II
can be expressed as
\begin{equation}
\label{cat}
\Gamma_{H^0({\rm or}h^0)} =
\frac{G_F}{4\sqrt2 \pi} M_{H^0} [3A\beta^3_c m^2_c
 + B(3\beta_b^3 m^2_b + \beta^3_{\tau}
m^2_{\tau})]
\end{equation}
where
\begin{equation}
\beta _f^2 = 1- \frac{4m_f^2}{m_\phi^2}    \hspace{2.5cm}   (\phi =H^0 \; or\;  h^0)
\end{equation}
In model I, $ A=({sin\alpha\over sin\beta})^2$, $B=({sin\alpha\over
sin\beta})^2$ for $H^0$, $A=({cos\alpha\over sin\beta})^2$, $B=({cos\alpha\over
sin\beta})^2$ for $h^0$, whereas in Model II,
$A=({sin\alpha\over sin\beta})^2$, $B=({cos\alpha\over
cos\beta})^2$ for $H^0$, $A=({cos\alpha\over sin\beta})^2$, $B=({sin\alpha\over
cos\beta})^2$ for $h^0$.

One alternative way to analyze the data is to measure the differential cross
section with respect to the invariant mass of $(p_b +p_{\overline{b}})^2
\equiv s_2$.
The interest is obvious: in the three body final state, the whole phase space
integration can smear out some information. Explicitly, if the $b$ and
$\overline{b}$ come from $b \overline{b} H^0$ and/or $b \overline{b} h^0$ vertices,
as $H^0(h^0)$
and the invariant mass of the $b\bar b$ pair crosses the pole, the
differential cross section can give rise to a peak.

Just as pointed out by ref.\cite{Boos}, an important feature is the large ratio
of the Higgs signal to the rest electro-weak background. For the
convenience of later
discussions, we give the explicit expression of
${d\sigma\over ds_2}$ for which only the Higgs, $H^0$ and $h^0$ contributions are
taken into account.
\begin{eqnarray}
\label{dog}
{d\sigma\over ds_2} &=& {3\over 128\pi^3s^2}({g^3m_b\over 8C_W^3})^2
(1+(1-4S^2_W)^2)
{1\over s_2}\lambda^{1/2}(s_2, m_b^2, m_b^2)\lambda^{1/2}(s, M_Z^2, s_2)\cdot
\nonumber\\
&& \{s+{1\over 4M_Z^2}[(s+M_Z^2-s_2)^2-{1\over 3}\lambda(s, M_Z^2,s_2)]\}\cdot
({s_2\over 2}-2m_b^2)\cdot
{1\over (s-M_Z^2)^2+\Gamma_Z^2M_Z^2}\nonumber \\
&& \times |{1\over (s_2-M^2_{H^0})
+i\Gamma_{H^0}M_{H^0}}{\rm cos}(\beta-\alpha)c_1 \nonumber \\
&& + {1\over (s_2-M^2_{h^0})
+i\Gamma_{h^0}M_{h^0}}{\rm sin}(\beta-\alpha)c_2|^2
\end{eqnarray}
where
$$\lambda(a,b,c)\equiv a^2+b^2+c^2-2ab-2bc-2ca$$
$S_W\equiv {\rm sin}\theta_W$, $C_W\equiv {\rm cos}\theta_W$,
$\theta_W$ is
the Weinberg angle, $c_1={\rm sin}\alpha/{\rm sin}\beta$, $c_2={\rm cos}
\alpha/{\rm sin}\beta$
for model I, $c_1=-{\rm cos}\alpha/{\rm cos}\beta$, $c_2={\rm sin}
\alpha/{\rm cos}\beta$  for Model II, and
$s=(p+p')^2$ with $p$ and $p'$ being the momenta of
electron and positron.

Boos and  Dubinin \cite{Boos} showed that the interference between the
"signal" diagram from Higgs and the other eight background diagrams is
small compared to itself of the signal and background at the peak of the 
Higgs resonance and the places far away from the peak respectively,
so one can investigate the signal of Higgs by  directly observing
the difference of the differential cross section from that predicted
by the well-understood background. The situation for the extended Standard
Model with two Higgs doublets is similar: the interference of the signal
caused by $H^0$ and $h^0$ with the eight background diagrams being small.
Therefore we can also study the change  of the cross section induced
by the two Higgs bosons in comparison with the background. Therefore
it makes sense that in
the figures for differential cross sections, we plot the total contributions
and the part from only the Higgs bosons separately.

Now, let us turn to the numerical analysis.\\

\section{The signal about Higgs and the background}

As aforementioned, if the MSM is valid, a precise measurement on the
cross section of $e^+e^-\rightarrow b\bar bZ^0$ would determine the Higgs
mass, but if the second Higgs doublet exists, it is more complicated and
uncertain.

The Feynman diagrams which concern $H^0$ and $h^0$ are only (3) and (4) of
Fig.1. Since all the four parameters $M_{H^0}, M_{h^0}, \alpha$ and $\beta$
are unknown, we cannot predict the cross section or differential cross section
precisely, instead, we will employ some specific values for the parameters
and clarify the physics picture. Moreover, there are constraints on the
$\beta-$value from the LEP experiments and
the $b\rightarrow s\gamma$ transition, namely very
small $\beta-$value ($tan\beta<0.2$) and very light $h^0$ ($M_{h^0}<60$ GeV)
regions are ruled out.

In fact, as the second Higgs boson $h^0$ is involved, the total cross section
and differential cross section would be different from that predicted by
the MSM. However our results show that the change of the total cross section
is too tiny for detecting.
The physically interesting observation is the differential cross section. 
We focus our attention on the the possibilities, which depend on the 
parameters of the model, i.e. due to possible but different parameters
of the model, at LEP II energy two peaks protruding 
out from the background occur in the differential cross section or only 
a single one does. We will show that even only one peak
exists in the $s_2$ sprectrum for LEP II experiments, the MSM and ESM 
still may predict a different width and height of the peak, hence their 
combined effect, the event rate, so different from each other that
the experiments may distinguish them, if the parameters in ESM are suitable.

(i) In Fig.2, we draw the differential cross section versus
$s_2\equiv (p_b+p_{\bar b})^2$ with $\alpha=\pi/4$, $M_{H^0}$=100 GeV,
$M_{h^0}$=70 GeV and $\beta=0.25$ in Model II, then it is found that as
$s_2$ varies, two resonance peaks appear very clearly above the background.

The upper curve which corresponds to the total contribution from all the ten
diagrams covers the lower one which only accounts for the two diagrams
concerning $H^0$ and $h^0$. It is noticed that the heights of the peaks
heavily depend on the parameter choices ($\alpha$ and $\beta$), but
the peak signal is obviously above the background and may be observable
if the resolution of the measurements on the momenta of $\bar b$ and $b$
is fine enough. The
widths also depend on the parameters, as shown in eq.(\ref{cat}).
The heights of the peaks are almost only determined by the Higgs
contribution. At the upper curve one can observe another broad peak at $M_Z$,
it is easy to understand that it comes from the $Z-$pole at (6) of Fig.1.

For Model I, the situation is very similar, we can clearly observe two peaks
at the $d\sigma/ds_2$ spectrum as for Model II, so for saving space, we
just omit it.

(ii) It would be interesting to investigate the possibility that if there is
only one peak in $s_2$ spectrum
over the possible energy range of LEP II, whether it corresponds to
and so confirms  the
contribution of the Higgs of the MSM or can be something else.
Besides the MSM, we may expect
another possible
solution. Namely, one of the peaks ($h^0$ or $H^0$) is located outside
our energy scan range, i.e. $M_{h^{0}\;{\rm or}\; H^0}>\sqrt s-M_Z$, so is
missing in the figures of differential cross sections.
Generally,
$M^{2HDM}_{h^0}<M^{2HDM}_{H^0}$, so we suppose that
the $H^0$ of the 2HDM is outside the
scan range.  Considering that the experimental resolution for 
measuring the momenta of
$\bar b$ anf $b$ pair is limited and the width of the Higgs resonance is
quite narrow, the quntities of $\Delta N$, 
where $\Delta N={d\sigma\over \sqrt s_2}\Gamma$,
relate to the event numbers directly and are not very sensitive to the 
experimental resolution, let us define the ratio:
\begin{equation}
R={(\Delta N^{2HDM})|_{h^0}\over (\Delta N^{MSM})|_{H^0}},
\end{equation}
and use the ratio $R$ to characterize the difference between 2HDM and
MSM. The superscript 2HDM and MSM
correspond to the Two-Higgs-Doublet-Model and the Minimal-Standard-Model
respectively. $R$ corresponds to the ratio of the events from
$h^0$ predicted by the 2HDM to that from $H^0$ by MSM, and
assuming $M^{MSM}_{H^0}=M^{2HDM}_{h^0}$ but $M^{2HDM}_{h^0}
\ll M_{H^0}^{2HDM}$ as well.

The dependence of $R$  on $\beta$ and $\alpha$ is shown
in Figs. 3 and 4 corresponding to Model I and Model II
respectively. The meaning of the results will be
discussed in next section.\\

\section{Discussions and conclusion}

Higgs hunting may be the task for the rest of this century, but we are
convinced by the past efforts in both experiments and theories that it is
a very difficult job. Any progress along the direction must be very exciting
and shed light on the mysterious sector of the Standard Model.

LEP II will run at $190 \sim 205$ GeV C. M. S energy and due to its much
clearer background than hadron collider, it is a
foreseeable ideal place for Higgs hunting
in the recent a few years.

In MSM, after SSB of $SU(2) \times U(1)$, only neutral Higgs remains. The
Higgs boson event rate for the bremsstrahling process $e^+ e^- \rightarrow
Z^0 H^0$ is known better than 1\% including radiative corrections
$\cite{Berends}$. To measure $e^+ e^-\rightarrow Z^0 b \overline{b}$ in fact is
measuring a combination of two processes $e^+e^-\rightarrow Z^0H^0$ and
$H^0\rightarrow b\bar b$, especially for the differential cross section at
the Higgs peak.
Namely if $H^0$ is not too heavy, the intermediate $H^0$
can be real. As the authors of ref.$\cite{Boos}$ showed that in the case
the ratio of signal
over background in $e^+ e^- \rightarrow b \overline{b} Z^0$ is greatly
increased and close to unity.
   
However, if the Higgs sector is not so simple, for example,
it includes two or several
doublets, the complexity increases. As we discussed above, analysis of
the total cross section depends on the employed theoretical models,
so a rash conclusion may be misleading.
If there indeed exists the second doublet in the Higgs sector, 
once we observe the total cross section
only, which is larger than that the supposed background can give
rise to, we still cannot be used it to relate to
the Higgs mass as done for MSM.
This makes the wish to draw a definite
conclusion on Higgs mass from measuring the total cross section of $e^+e^-
\rightarrow b\bar bZ^0$ pessimistic.

As shown in Fig.2, the differential cross section with
$\frac{d\sigma}{d(p_b+p_{\overline{b}})^2}\equiv {d\sigma\over ds_2}$
indeed demonstrates two peaks which correspond to $H^0$ and $h^0$
respectively because $H^0$ and $h^0$ both are not too heavy. It
certainly is an evidence of ESM.
Of course, there is another possibility that $M_{H^0}$ is too
large that its peak cannot be allowed to appear
by the phase space for the LEP II energy.
In this possible case, one probably observe one peak only,
but a careful analysis of the measurements,
which include the total cross section and differential one,
may still help to distingruish it from $h^0$ of 2HDM or the Higgs boson of
the MSM. 

In general, it is more interesting to study the case when only one 
peak exists in the figures for the differential cross section at
a precise energy e.g. at LEP II, because in the case the feature 
of the signal is similar for MSM and 2HDM, whereas it is still possible 
to indicate whehter the peak corresponds to MSM or 2HDM in certain
conditions. In the paper, we would like to see the conditions:
the experimental measurements are precise enough and the parameters of 2HDM
are suitable. In Figs.3,4 we demonstrate the dependence of $R$  
defined in last section on the two angles $\beta$ and $\alpha$ of 2HDM.
Thus from the Figs,2-4 we may achieve some understanding of the peak when
having the peak well measured.
If $R\sim 1.0$, one would not be able to distingruish MSM and 2HDM, while
if one may be certain to exclude the uncertainties and
to have $R\leq 1.0$ (from Figs.3,4 one may see that there are
very rare chances for the model parameters to have $R > 1.0$), 
one would be able to say the peak is in favor of 2HDM
with a possible choice of the model parameters.

In fact, in terms of eq.(10), one can immediately obtain  approximate
expressions of $R$ for Model I and II. Concretely,
\begin{eqnarray}
R &\approx & \sin^2(\beta-\alpha) \hspace{7cm}({\rm for\; Model\; I})\\
R &\approx & \sin^2(\beta-\alpha){1\over 0.8 \cot^2\alpha\cot^2\beta+0.92}
\hspace{3cm} ({\rm for\; Model \; II}).
\end{eqnarray}
The results show that if one resonance of $H^0$ or
$h^0$ (usually assuming $M_{h^0}<M_{H^0}$) is outside our energy scan range
($M_{H^0}>\sqrt s-M_Z$ in our case), for Model I of 2HDM,
with a reasonable $\beta$ range as
$tan\beta>0.21$, which is constrained by the data of $b\rightarrow s\gamma$
\cite{stange}\cite{hewett}, we always have $R<1$, namely the area encompassed
by the peak resulted by the neutral Higgs of the 2HDM is always smaller
than that resulted by the MSM Higgs. Whereas, for Model II, 
there is possibility that $R>1$, but within a plausible region
$0.21<\beta<\pi/2$ and $\alpha$ not being large, 
the ratio $R$ is also smaller than unity.

LEP II will provide an integrated luminosity of about 170 $pb^{-1}$ per year
\cite{Augustin}
with the data taking efficiency less than 25 \%. According to
the estimation of ref.\cite{Boos}, the MSM can produce less than 
100 events every year as LEP II operates at 195 GeV. With the number 
as a reference, our results indicate that the total cross section
and differential cross section can be observed when $R$ is not too small
e.g. $R\geq 0.1$, but still vary with the Higgs mass in models.

To determine if the peak corresponds to MSM or 2HDM, one
should require $R\leq 0.7$, otherwise a clear judgement is very hard
if not impossible at LEP II due to very rare events.
From our numerical results shown in Fig.3 and Fig.4, only $\alpha$ and
$\beta$ remain within certain ranges, $R$ can be expected to be less than
$0.7$ but greater than $0.1$. As discussed, the complexity due to the 
two Higgs doublets cannot be eliminated by the unique process 
$e^+e^-\rightarrow b\bar bZ^0$ at LEP II. In fact, it is very limited for 
LEP II to solve the problem. As the physical world sets an even messier 
picture to us, along the direction any progress will be inspiring and 
encouraging and the measurements on $e^+e^-\rightarrow
b\bar bZ^0$ are definitely significant in the Higgs hunting process
\cite{Diaz}\cite{Chang}\cite{Boos1}.

Our conclusion is that even though the $e^+e^- \rightarrow b\overline{b} Z^0 $
measurements at LEP II can provide us some direct evidence and 
information about $H^0$ to indicate MSM or 2HDM if we are so lucky enough
that the Higgs mass eventually falls into the experimental ability, 
the new scenario will begin immediatelly in fact. 
To determine the Higgs doublet structure is a complicated and very hard
problem with such a few events. A careful measurement on the 
differential cross section $\frac{d\sigma}{d(p_b+p_{\overline{b}})^2}$
is always useful and helpful, especially for determining the mass
of the Higgs.

\section*{Acknowledgements}

This work was supported in part by the National Natuare Science Foundation
Of China and the Grant LWTZ-1298 of Chinese Academy of Sciences.\\

\vspace{1cm}

\noindent {\large Figure Captions}\\

Fig.1. The Feynman diagrams for $e^+e^-\rightarrow b\bar bZ^0$, where
(3) and (4) are that concern $H^0$ and $h^0$ respectively.\\

Fig.2. The dependence of the differential cross section on $s_2$ for Model II
with $\alpha=\pi/4$, $\tan\beta=0.25$,
$M_{H^0}=$100 GeV and $M_{h^0}=$70 GeV.\\

Fig.3 The dependence of $R$ on $\alpha$ and $\beta$ for model I.\\

Fig.4 The dependence of $R$ on $\alpha$ and $\beta$ for model II.

\end{document}